# Transparent EuTiO$_3$ films: a novel two-dimensional magneto-optical device for light modulation

Annette Bussmann-Holder[1*], Krystian Roleder[2], Benjamin Stuhlhofer[1], Gennady Logvenov[1], Iwona Lazar[2], Andrzej Soszyński[2], Janusz Koperski[2], Arndt Simon[1], Jürgen Köhler[1]

[1]Max-Planck-Institut für Festkörperforschung, Heisenbergstr. 1, D-70569 Stuttgart, Germany

[2]Institute of Physics, University of Silesia, ul. Uniwersytecka 4, 40-007 Katowice, Poland

The magneto-optical activity of high quality transparent thin films of insulating EuTiO$_3$ (ETO) deposited on a thin SrTiO$_3$ (STO) substrate with both being *non-magnetic* materials are demonstrated to be a versatile tool for light modulation. The operating temperature is close to room temperature and admits multiple device engineering. By using small magnetic fields birefringence of the samples can be switched off and on. Similarly, rotation of the sample in the field can modify its birefringence Δn. In addition, Δn can be increased by a factor of 4 in very modest fields with simultaneously enhancing the operating temperature by almost 100K.

ETO has the cubic perovskite structure at room temperature [1] and undergoes a structural phase transition to tetragonal at $T_S$=282K [2]. Below $T_N$=5.7 K the Eu 4f$^7$ spins order G-type antiferromagnetic [3], and large magneto-electric coupling takes place as evidenced by the magnetic field dependence of the dielectric constant [4]. Very unusual magnetic field dependent properties are observed in the paramagnetic phase at high temperature as demonstrated by the field dependence of $T_S$ [5] and anomalies in the magnetic susceptibility at $T_S$ [6]. These results indicate some kind of *hidden magnetism* in the paramagnetic phase which is supported by muon spin rotation (μSR) data where a strong field dependence of the μSR relaxation rate is observed [7]. Similarly, resonant ultrasound spectroscopy (RUS) experiments [8] reveal a pronounced influence of a magnetic field on the acoustic properties of ETO.

**Results**

Bulk samples of ETO are available only as tiny single crystal or in ceramic form and exhibit large leakage currents which make them unsuited for optical measurements. Thin films of ETO have been fabricated by various groups [9-13], however, all being unable to overcome the leakage problem. Here we report on results from films of ETO deposited on a thin STO substrate which are highly transparent, single crystalline, cubic at room temperature, and strain/stress free.

These films allowed to observe for the first time their birefringence properties in the tetragonal phase and enabled to detect a further structural phase transition at T*≈190K from tetragonal to monoclinic [14]. While T* has already been identified by µSR [7] as a crossover temperature below which some kind of magnetic order appears, it is evident from the present data that a real phase transition takes place at T* with novel unexpected properties.

Details of the sample preparation and their characterisation are given in the supplementary material section A. The samples have been confirmed to be antiferromagnetic below $T_N$=5.1K and to undergo the cubic – tetragonal transition at $T_S$=282K (Figure SM 1). The investigated films thus display all properties as ceramic bulk and single crystal samples, however, being superior to those due to the avoidance of the leakage problem.

As is obvious from Figure SM1, besides of $T_S$=282K a second phase transition occurs around T*≈190K as indicated by a change of the slope of the temperature dependence of Δn. This is shown more clearly in the inset to Figure SM1 where the Landau type behavior (straight line of the main figure) has been subtracted from the data. In order to explore the properties of ETO around and below T* and derive the symmetry of the structure, the birefringence data were taken in a magnetic field with the field direction being rotated such that H was parallel to [100], [010], [110], [1$\bar{1}$0], respectively. For this purpose the Metripol Birefringence Imaging System (Oxford Cryosystems) [15] has been modified to admit the application of a magnetic field. Further details about the birefringence measurements and the Metripol system are given in the supplementary material SM B. The sample's orientation was carefully checked with the tetragonal c-axis being well oriented in [001] direction of the STO substrate. Three magnetic field strengths have been used, namely H=0.02T, 0.063T and 0.1T. An overview over all orientations mentioned above in the smallest field of 0.02T is given in Figure 1 (left) in the temperature range between 100 and 340K. At a first glance the complete loss of birefringence with the field along the [110] direction for T<$T_S$ and its abrupt onset at T* is very striking. This is in stark contrast to the data taken in the opposite direction, namely along [1$\bar{1}$0] where Δn smoothly increases below ~240K to steeply increase around 180K. Along the [100] and [010] directions the behavior of Δn is very different since the birefringence becomes finite at $T_S$ and increases linearly, as expected from Landau theory, and exhibits a maximum around T* followed by an almost temperature independent regime for T<T*. Since tiny misalignments of the sample with respect to the external field cannot be completely ruled out, the data in Figure 1a have been normalized to their maxima values in the displayed temperature region and are shown in Figure 1b.

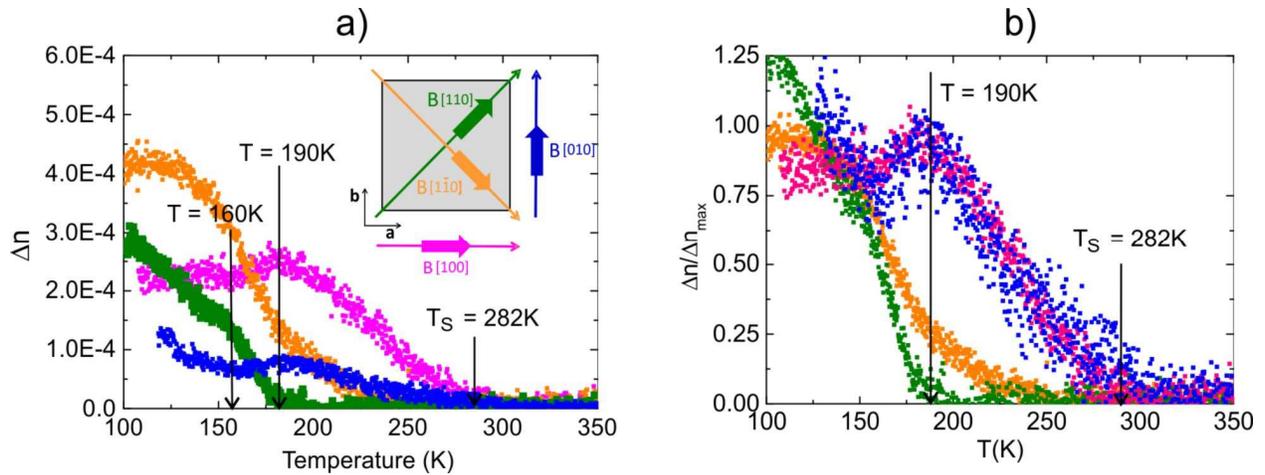

**Figure 1 (a)** Birefringence of the ETO film as a function of temperature in a magnetic field of 0.02T. The magnetic field orientation with respect to the crystallographic axis of the sample is shown in the inset to the figure with the colour code corresponding to the respective data. The horizontal lines indicate $T_S$, $T^*$ and the onset of precursors of STO (see supplementary material section B.) **(b)** The same as in the left hand picture, however, with the data being normalized with respect to their maximum around 180K for the blue and pink data and around 100K for the orange and green data. The color code for the field directions is the same as the one in Figure 1a. The horizontal lines mark $T_S$ and $T^*$.

The distinct differences between all field directions are now less pronounced and especially the data taken along [100] and [010] are almost identical, at least in the range between $T^*$ and $T_S$ where the tetragonal symmetry is realized. However, [110] and [1$\bar{1}$0] differ substantially from those and between each other already well above $T^*$. The zero birefringence in both directions below $T_S$ is well understandable since in these directions the tetragonal domain structure renders them isotropic. The differences between both appearing below 250K evidence, however, that the structure has changed with the most dramatic change setting in at $T^*$. A visualization of these alterations is best achieved by looking at the orientation images, i.e., the angle $\Phi$ changes $\Delta\Phi$ of the inclination of the optical indicatrix, as obtained for all directions and selected temperatures (Figures 2).

In Figures 2 the magnetic field is oriented along $[1\bar{1}0], [110], [010],$ and $[010]$, from a) to d), respectively. Blatant differences are well apparent in the images with the most striking features appearing along the diagonal field orientation $[1\bar{1}0]$. At T=280K the image corresponds to the tetragonal phase where $\Delta n=0$, analogous to what is observed in the opposite direction. At T*=190K sharp domain structures have developed resembling a checkerboard pattern which increase in intensity upon lowering the temperature. This is in contrast to what is observed in the opposite direction where domains become visible in terms of stripes again gaining intensity with decreasing temperature. The directions along the main axes [100] and [010] are different since here a finite phase difference is obvious at T=240K as expected in the tetragonal phase. The complicated domain pattern as seen along $[1\bar{1}0]$, is, however, absent and more stripe like features appear with decreasing temperature.

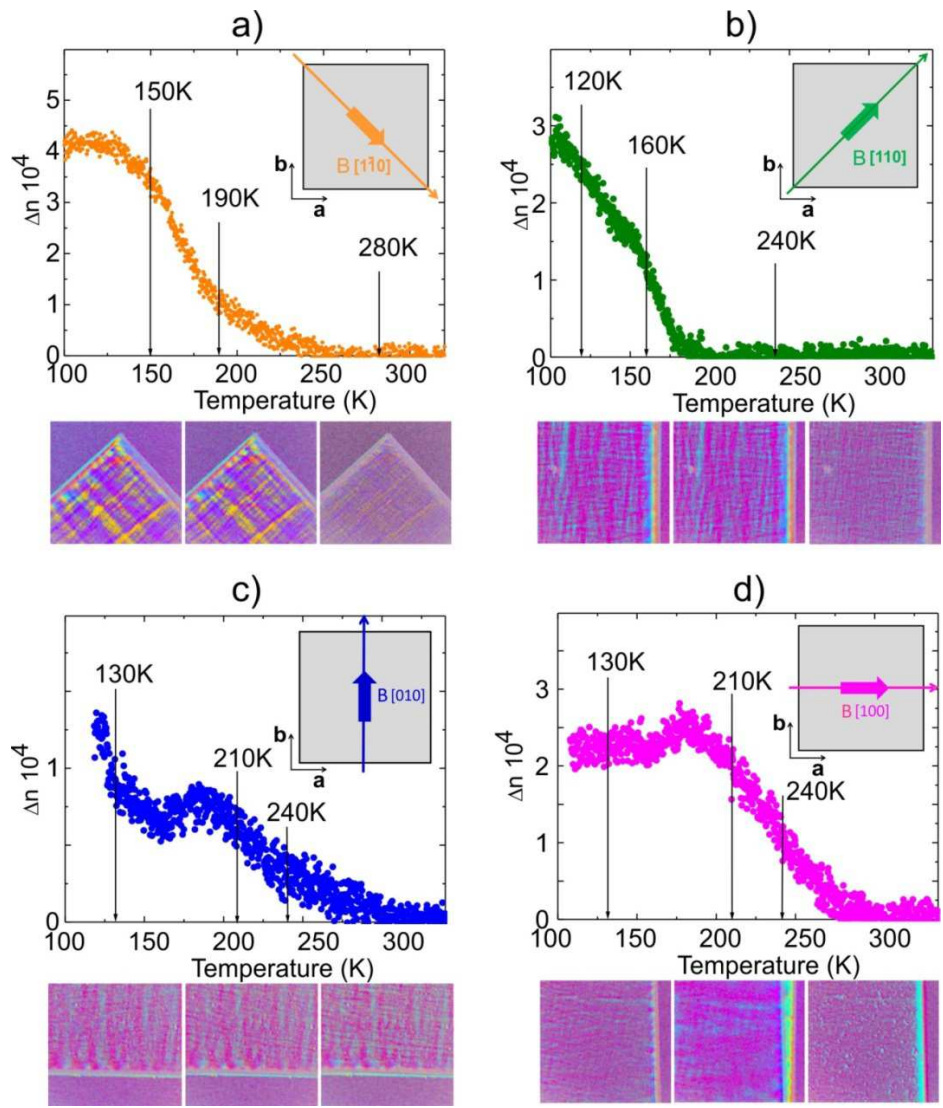

**Figure 2** a) to d) Temperature dependence of the birefringence (upper parts to the figures) in a magnetic field of H=0.02T oriented along (a) $[1\bar{1}0]$, (b) [110], (c) [010], and (d) [100]. The horizontal lines indicate the temperatures at which the orientation images of the birefringence (lower parts to the figures) have been taken. It is important to note that birefringence reveals also quite regular but different stresses at each edge of the sample.

Since the data presented above are visible only in an external magnetic field, we have increased the field strength from 0.02T to 0.1T, still being moderately small for possible device designs. The results are shown in Figures 3.

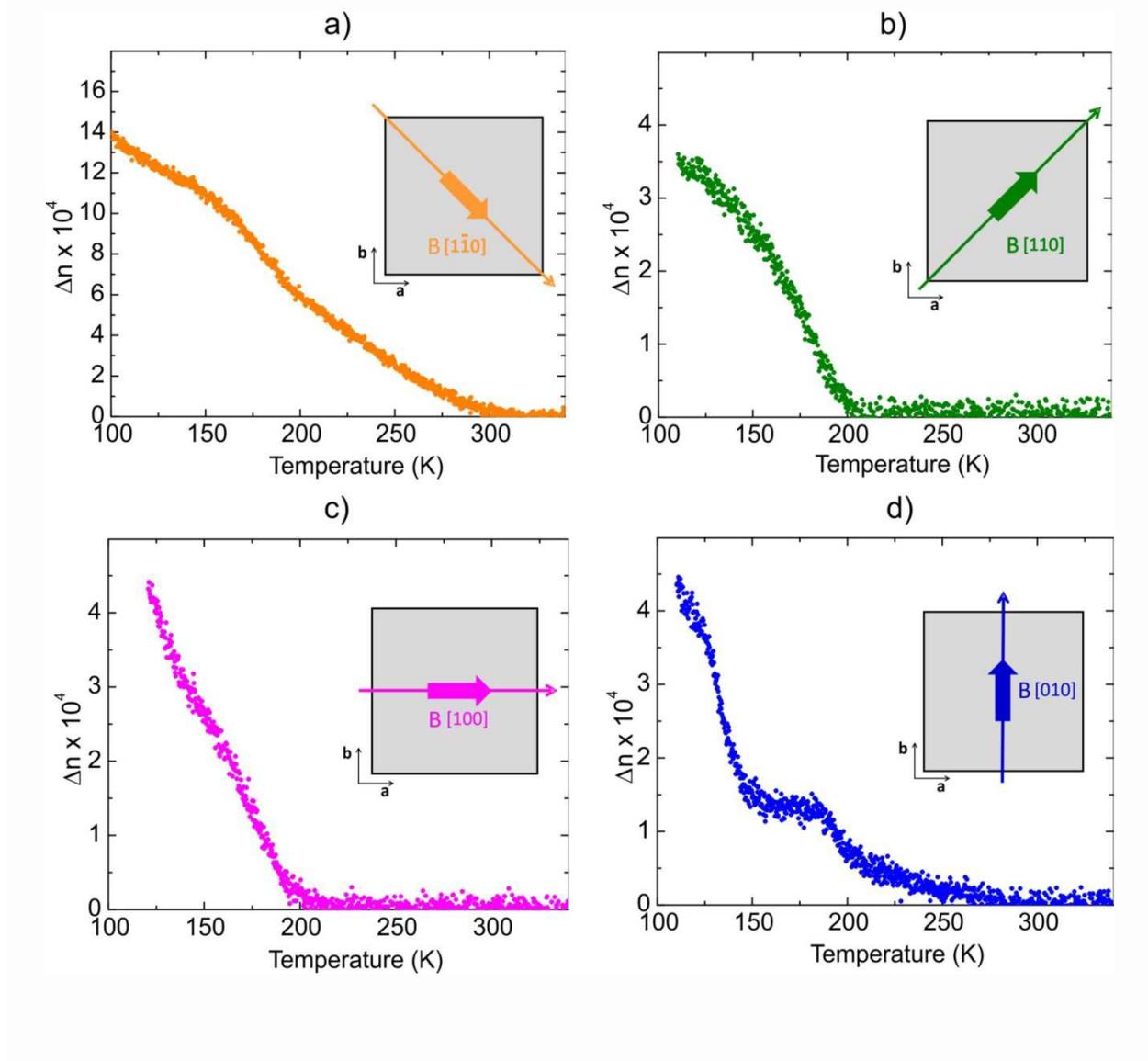

**Figure 3** Birefringence of the ETO film in a magnetic field of H=0.1T with the field being directed along **(a)** $[1\bar{1}0]$, **(b)** [110], **(c)** [100], and **(d)** [010].

Before discussing the data it is important to note that the scale of the birefringence between Figure 3a and the remaining figures has been changed by a factor of ~3. Now very pronounced differences appear along all directions, where only Figure 3a and 3d reveal distinct signatures of the phase transition from cubic to tetragonal at $T_S$=282K, quite opposite to the low field data. The transition at T* is clearly visible in Figures 3b and 3c signalled by a marked sudden onset of Δn. However, also along [010] this transition appears in the form of a maximum followed by a further increase in Δn below 150K while only a small anomaly distinguishes this transition along $[1\bar{1}0]$. From these results it must be concluded that a magnetic field of only 0.1T dramatically influences Δn and the transition at T* and induces another phase transition at $T_S$=282K with the symmetry being different from tetragonal since Δn differs substantially along the directions [100] and [010] incompatible with tetragonal symmetry. On the other hand [010] and [110] are rather similar supporting this conclussion. The most striking feature in figures 3 is, however, the huge increase in Δn along $[1\bar{1}0]$, where it is more than three times larger than without field or with the field being H=0.02T. This opens avenues for device designs by tuning the transparency of the films by a magnetic field. Since the data for H along $[1\bar{1}0]$ are the most striking ones and exhibit the most dramatic effects, their ΔΦ images are shown in Figure 4a-c where also the transition at T=282K is distinctly recognizable since at T=240K < $T_S$ clear signs of a finite Δn are obvious.

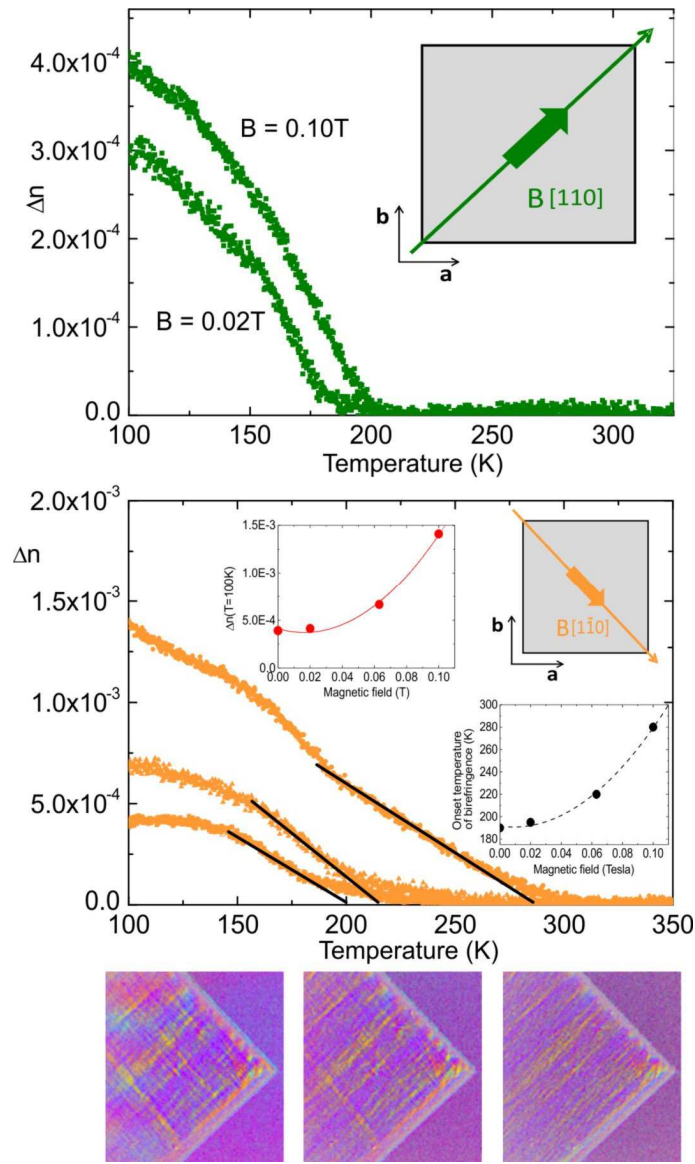

**Figure 4 (top to bottom) (top)** Birefringence of the film in a field of H=0.02T (circles), and H=0.1T (squares) with the field direction along [110]. **(middle)** The same as shown in (top) with the field along [1$\bar{1}$0], and field strengths H=0.02T (circles), 0.063T (triangles), and H=0.1T (open squares). The straight lines refer to the extrapolated onset temperatures of Δn as shown in the inset (right) to this figure. The insets show the onset temperature of Δn versus magnetic field (right) and Δn(T=100K) versus H (left). **(bottom)** ΔΦ images in a magnetic field of H=0.1T taken along [1$\bar{1}$0] for temperatures T=85, 170, 240K (from left to right).

While for $T_S$>T>T* stripe like domains appear, these change to a checkerboard pattern below T* which stays down to temperatures as low as 85K with its brightness increasing. The change from stripes to checkerboards goes hand in hand with the transition at T*. Since the two orthogonal directions [110] and [1$\bar{1}$0] show the most amazing behavior in the magnetic field, the data for both are compared to each other in Figure 4(top) and 4(middle) for the two field strengths of H=0.02T and 0.1T. Note again that the scale of the y-axis has been changed between both figures, with the one in 4(bottom) being more than three times larger than in 4(top). In both cases it is, however, seen that the magnetic field shifts the onset of Δn to higher temperatures and simultaneously increases Δn. This effect can be quantified for the [1$\bar{1}$0] direction where more complete data on the magnetic field effects are available. In the insets to Figure 4(middle) the data for H=0, 0.02, 0.063 and 0.1T are summarized. Both, the transition to the non-birefringent state as well as Δn increase nonlinearly with the field strength which demonstrates the enormous sensitivity of the Δn with respect to an external magnetic field. The very pronounced differences in the data along the nominally orthogonal directions [1$\bar{1}$0] and [110] highlight the inequivalence between them which – as long as the system is in the tetragonal phase – should not be the case. In order to observe such a discrepancy in these two directions, the angles between [100]/[110] and [100]/[1$\bar{1}$0] must be unequal as is realized in the monoclinic structure, not in the orthorhombic symmetry which would be the typical symmetry lowering sequence in perovskites. Assuming that this assignement is correct it renders the Eu ions within the [100] basal plane inequivalent being incompatible with the cubic and the tetragonal phase whereas possible in the orthorhombic symmetry. The latter, however, does not permit the diagonals to be inequivalent. In the monoclinic structure the additional deviation of the planar angles from 90° has the effect that two Eu ions approach each other along the plane diagonal whereas the other two move away from each other (For details see SMC). Since we have recently shown that the transition to antiferromagnetic order at low temperature is accompanied by tiny but significant contractions of the lattice [15], demonstrating that the Eu-Eu distance is decisive for the spin alignement, similar effects must be the origin of the enormous sensitivity of ETO to small magnetic fields. Increasing the field introduces an additional symmetry lowering already in the tetragonaal phase since [100] and [010] differ considerably from each other. The additonal field effect on the onset temperatures for the birefringence suggests that the Eu spins assume a certain magnetic order which triggers the symmetry lowering. The complex interplay between spin dynamics, structural phase transitions and magnetic field thus offers an great potential for novel applications.

**Discussion**

With respect to possible device designs of ETO, we suggest to implement it in a two-dimensional magneto-optical light modulator for signal processing. For this purpose the ETO film should be structured into isolated mesas, placed in a magnetic field and cooled to temperatures below $T_S$, i.e., close to room temperature. Depending on the orientation direction of the mesas, bright or dark spots appear which – in turn – can be reoriented to work purpose adapted. The operating temperatures are well accessible and the switching speed is fast with high stability. Another option for light modulation functions is the use of a rotating magnetic field or conversely to place the film on a rotating disk. With varying rotation angle, light is either transmitted or not. Another more sophisticated aplication is the detection of a magnetic field by the change of the birefringence, especially if the magnetic field is directed along the $[1\bar{1}0]$ direction. In this case moderately small magnetic fields transform the ETO films into highly anisotropic optical materials. Many more applications of these films can be thought of and further work is in progress.

**Conclusions**

Thin films of ETO deposited on an STO substrate have been fabricated and investigated by birefringence which was only possible due to the superior quality of the films evident from their high transparency. Besides of the well known structural phase transition at $T_S$=282K and the low temperature transition to antiferromagnetic order at $T_N$=5.1K a novel phase transition has been discovered at T*=190K. This transition to monoclinic symmetry can be influenced by small magnetic fields directed along the main symmetry directions of the sample, namely [100], [010], [110], and [1$\bar{1}$0], respectively. The birefringence along these directions changes dramatically with the field and can be tuned over 100K. A pronounced domain pattern formation is accompanied with it which can be checker board or stripe like. The possibility to tune the birefringence offers new device applications in terms of two-dimensional magneto-optical light or more complex spatial light modulators. The important issue is, however, that the observed effects take place in a compound which is nominally not magnetic including the substrate and sufficiently thick to exclude interfacial phenomena to be responsible for them.

magnetic field measurements, B. S., G. L. have fabricated the thin films, A. S., J. K. have prepared the target material.

Additional Information

**Competing financial interests:** The authors declare no competing financial interests.